\documentclass[11pt,fleqn,letterpaper]{article}

\usepackage{cite}
\usepackage{amsmath,amssymb,amsfonts}
\usepackage{algorithmic}
\usepackage{graphicx}
\usepackage{textcomp}
\usepackage{xcolor}
\usepackage{url}
\usepackage{verbatim}
\usepackage{multirow}
\usepackage{comment}
\usepackage[title]{appendix}

\title{User-Driven Abstraction for Model Checking}
\author{Glenn Bruns\\
        Bell Labs, Lucent Technologies}
\date{April, 1998}

\begin{document}  
\bibliographystyle{plain}         
\maketitle

% basic types
\newcommand{\Names}{\mbox{$\cal A$}}
\newcommand{\Conames}{\mbox{$\overline{\Names}$}}
\newcommand{\Labels}{\mbox{$\cal L$}}
\newcommand{\VPLabels}{\Labels_{vp}}
\newcommand{\Actions}{{Act}}
\newcommand{\AgentExprs}{\mbox{$\cal E$}}
\newcommand{\Agents}{\mbox{$\cal P$}}
\newcommand{\Proc}{{\cal P}}
\newcommand{\States}{\mbox{${\cal P}$}}

% constants and actions
\newcommand{\Ct}[1]{\mbox{\it #1\/}}
\newcommand{\Na}[1]{\mathop{\tt #1}\nolimits}
\newcommand{\Co}[1]{\mathop{\overline{\tt #1}}\nolimits}

\newcommand{\eps}{\varepsilon}

% process operators
\newcommand{\nil}{{\bf 0}}
\newcommand{\prefix}[2]{\mbox{$#1 . #2$}}
\newcommand{\choice}[2]{\mbox{$#1 + #2$}}
\newcommand{\comp}[2]{\mbox{$#1 \mid #2$}}
\newcommand{\relabel}[2]{#1 [ #2 ]}
\newcommand{\restrict}[2]{\mbox{$#1 \setminus #2$}}
\newcommand{\rest}[1]{\mbox{$\setminus #1$}}
\newcommand{\res}{\mbox{$\setminus$}}
\newcommand{\hide}[1]{\backslash\!\!\backslash #1}
\newcommand{\hid}{\mbox{$\backslash\!\!\backslash$}}
\newcommand{\Def} {\stackrel{{\rm def}}{=}}
\newcommand{\Empty}{\mbox{$\cal E$}}
\newcommand{\Sort}[1]{\mbox{${\cal L}(#1)$}}
\newcommand{\Rch}[1]{{\rm Rch}(#1)}
\newcommand{\Pair}[2]{#1 \! / \! #2}
\newcommand{\RePair}[2]{\Na{#1} \! / \! \Na{#2}}

% process relations
\newcommand{\Eq}[2]{\mbox{$#1 \sim #2$}}
\newcommand{\ObsEq}[2]{\mbox{$#1 \approx #2$}}

% families of definitions
\newcommand{\Fam}[1]{{\cal #1}}
\newcommand{\ProcRel}[2]{#1 \mbox{ of } #2}
\newcommand{\Merge}[3]{{\rm merge}(#1,#2,#3)}

% value-passing
\newcommand{\If}{\mbox{\hspace{1pt}if }}
\newcommand{\Then}{\mbox{ then }}
\newcommand{\Else}{\mbox{ else }}
\newcommand{\Let}{\mbox{let }}
\newcommand{\In}{\mbox{ in }}
\newcommand{\til}{\mbox{$\tilde{\ }$}}
\newcommand{\col}{\mbox{$\!\, :  \!\,$}}

% transition relations
\newcommand{\arrow}[1]  {\mbox{$\stackrel{#1}{\rightarrow}$}}
\newcommand{\larrow}[1] {\mbox{$\stackrel{#1}{\longrightarrow}$}}
\newcommand{\warrow}[1] {\mbox{$\stackrel{#1}{\Rightarrow}$}}
\newcommand{\lwarrow}[1]{\mbox{$\stackrel{#1}{\Longrightarrow}$}}
\newcommand{\nonarrow}[1]{\mbox{$\not \stackrel{#1}{\rightarrow}$}}
\newcommand{\OpArrow}[1]{\mbox{${\stackrel{#1}{\longrightarrow}}_{\cal O}$}}
\newcommand{\trans}[3] {#1 \, \arrow{#2}  \, #3}
\newcommand{\ltrans}[3]{#1 \, \larrow{#2} \, #3}
\newcommand{\wtrans}[3]{#1 \, \warrow{#2} \, #3}
\newcommand{\nontrans}[1]{\mbox{$#1 \not{\hspace{-3pt}\rightarrow\hspace{1pt}}$}}
\newcommand{\PLabels}{\mbox{${\cal L}_p$}}
\newcommand{\PActions}{\mbox{$Act_p$}}
\newcommand{\PProc}{{\cal P}_p}
\newcommand{\parrow}[2]{{\stackrel{#1:#2}{\longrightarrow_p}}}
\newcommand{\pnonarrow}[2]{{\not \stackrel{#1:#2}{\longrightarrow}}_p}
\newcommand{\ptrans}[4]{#1 \, {\stackrel{#2:#3}{\longrightarrow_p}} \, #4}
\newcommand{\panytrans}[1]{#1 \rightarrow_p}
\newcommand{\pnontrans}[1]{#1 \not{\hspace{-3pt}\longrightarrow_p\hspace{1pt}}}
\newcommand{\Pa}[2]{#1 \hspace{-2pt} : \hspace{-2pt} #2}

% modal process logic
\newcommand{\MSpecs}{\mbox{$\cal S$}}
\newcommand{\MustArrow}[1]{\mbox{${\stackrel{#1}{\longrightarrow}}_{\Box}$}}
\newcommand{\MayArrow}[1] {\mbox{${\stackrel{#1}{\longrightarrow}}_{\Diamond}$}}
\newcommand{\MArrow}[1]   {\mbox{${\stackrel{#1}{\longrightarrow}}_{m}$}}
\newcommand{\MustTrans}[3]{#1 \MustArrow{#2} #3}
\newcommand{\MayTrans}[3] {#1 \MayArrow{#2} #3}
\newcommand{\MTrans}[3]   {#1 \MArrow{#2} #3}
\newcommand{\MustAct}[1]  {#1_{\Box}}
\newcommand{\MayAct}[1]   {#1_{\Diamond}}
\newcommand{\Refine}[2]   {#1 \lhd #2}

\newcommand{\WMustArrow}[1]{\mbox{$\stackrel{#1}{{\Longrightarrow}}_{\Box}$}}
\newcommand{\WMayArrow}[1] {\mbox{$\stackrel{#1}{{\Longrightarrow}}_{\Diamond}$}}
\newcommand{\WMArrow}[1]   {\mbox{$\stackrel{#1}{{\Longrightarrow}}_{m}$}}
\newcommand{\WMustTrans}[3]{\mbox{$#1 \WMustArrow{#2} #3$}}
\newcommand{\WMayTrans}[3] {\mbox{$#1 \WMayArrow{#2} #3$}}
\newcommand{\WMTrans}[3]   {\mbox{$#1 \WMArrow{#2} #3$}}
\newcommand{\WeakRefine}[2]{\mbox{$#1 \unlhd #2$}}

% miscellany
\newcommand{\Exp}[2]      {\mbox{$#1 \,\leq\, #2$}}
\newcommand{\ERef}[2]     {\mbox{$#1 \,\preceq\, #2$}}
\newcommand{\update}[2]{\mbox{$[ #1 / #2 ]$}}

% logics
\newcommand{\Mu}{\mbox{$\mu L$}}
\newcommand{\Mup}{\mbox{$\mu L^+$}}
\newcommand{\Muw}{\mbox{$\mu I \!\! L$}}
\newcommand{\Mub}{\mbox{$\mu L \Box$}}
\newcommand{\Mupw}{\mbox{$\mu I\!\!L^+$}}
\newcommand{\Mubw}{\mbox{$\mu I\!\!L\Box$}}
\newcommand{\Mubc}[1]{\mbox{$\mu L \Box_{#1}$}}
\newcommand{\Mur}{\mbox{$\mu Lr$}}

% formula names
\newcommand{\Prop}[1]{\mbox{\it #1\/}}

% propositional operators
\newcommand{\Tr}{\mathop{{\tt tt}}\nolimits}
\newcommand{\Fa}{\mathop{{\tt ff}}\nolimits}
\newcommand{\IffArrow}{\Leftrightarrow} 
\newcommand{\Equiv}{\leftrightarrow}
\newcommand{\MyAnd}{\,\wedge\,} 
\newcommand{\Or}{\,\vee\,}
\newcommand{\Not}{\neg}
\newcommand{\Implies}{\Rightarrow}
\newcommand{\ImpliedBy}{\Leftarrow}
\newcommand{\Impl}{\rightarrow}

\newcommand{\PNot}{\mbox{not }}
\newcommand{\POr}{\mbox{ or }}
\newcommand{\PAnd}{\mbox{ and }}

% modal operators
\newcommand{\may}[1] {\mbox{$ \langle #1 \rangle $}}
\newcommand{\must}[1]{[ #1 ]}
\newcommand{\wmay}[1] {\mbox{$ \langle \! \langle #1 \rangle \! \rangle $}}
\newcommand{\wmust}[1]{\mbox{$ [ \! [ #1 ] \!  ] $}}
\newcommand{\May}[2] {\mbox{$ \langle #1 \rangle #2 $}}
\newcommand{\Must}[2]{\mbox{$ [ #1 ] #2 $}}
\newcommand{\Wmay}[2] {\mbox{$ \langle \! \langle #1 \rangle \! \rangle #2 $}}
\newcommand{\Wmust}[2]{\mbox{$ [ \! [ #1 ] \!  ] #2 $}}
\newcommand{\Kmay}[1] {\mbox{$ \langle #1 \rangle^{\ast} $}}
\newcommand{\Kmust}[1]{\mbox{$ [ #1 ]^{\ast} $}}
\newcommand{\Pmay}[3] {\mbox{$ \langle \! \langle #1 (#2) \rangle \! \rangle #3 $}}
\newcommand{\Pmust}[3]{\mbox{$ [ \! [ #1(#2) ] \!  ] #3 $}}

% valuations
\newcommand{\Val}{{\cal V}}
\newcommand{\EmptyVal}{{()}}
\newcommand{\UpdateVal}[3]{#1[#2 := #3]}

% semantics
\newcommand{\TrSys}{{\cal T}}
\newcommand{\interp}[1]{\mbox{$\| #1 \| $}}
\newcommand{\Interp}[1]{\interp{#1}^\TrSys_\Val}
\newcommand{\setprop}[1]{\mbox{$ \interp{#1}_{\cal V} $}}
\newcommand{\Sat}{\models} 
\newcommand{\decls}{\bigtriangleup}

\newtheorem{proposition}{Proposition}
\newtheorem{observation}{Observation}
\newtheorem{conjecture}{Conjecture}
\newtheorem{remark}{Remark}
\newtheorem{theorem}{Theorem}
\newtheorem{lemma}{Lemma}
\newtheorem{definition}{Definition}
\newtheorem{corollary}{Corollary}

\newcommand{\Dom}     {\mbox{Dom}}
\newcommand{\func}[2] {\mbox{$ \lambda #1 . #2 $}}
\newcommand{\Fin}[2]  {\mbox{$#1 \stackrel{\mathrm fin}{\rightarrow} #2$}}
\newcommand{\Seqs}[1] {\mbox{\mathrm Seq}(#1)}
\newcommand{\Nat}     {{I \! \! N}}
\newcommand{\Real}    {{Z \!\!\! Z}}
\newcommand{\Rational}{{\, l \!\!\! Q}}
\newcommand{\rsymbol} {{I \! \! R}}

\newcommand{\BnfBar}  {\,\mathrel{\vrule width 0.7pt height 1.1em depth 0.3em}\,}
\newcommand{\Proof}   {\par\noindent{\bf Proof.\ }}
\newcommand{\QED}     {\quad\hfill\mbox{$\Box$}\par\medskip}
\newcommand{\abst}[1] {\mbox{${\cal A}_f(#1)$}}
\newcommand{\fsp}     {\hspace*{.4 in}}
\newcommand{\hsp}     {\hspace*{.1 in}}
\newcommand{\denot}[1]{[\![ #1 ]\!]}
\newcommand{\tdel}[1] {\mbox{$<\!< #1 >\!>$}}
\newcommand{\semd}[1] {\mbox{$ \cal{D} \denot{#1} $}}
\newcommand{\seme}[1] {\mbox{$ \cal{E} \denot{#1} $}}
\newcommand{\semc}[1] {\mbox{$ \cal{C} \denot{#1} $}}
\newcommand{\seml}[1] {\mbox{$ \cal{L} \denot{#1} $}}
\newcommand{\EP}[3]   {\mbox{$ #1 \vdash #2 \Rightarrow #3 $}}
\newcommand{\ET}[4]   {\mbox{$ #1 \vdash \trans{#2}{#3}{#4} $}}
\newcommand{\EB}[4]   {\mbox{$ #1 \vdash_b \trans{#2}{#3}{#4} $}}
\newcommand{\EA}[5]   {\mbox{$ \ltrans{(#1,#2)}{#3}{(#4,#5)} $}}
\newcommand{\Opt}[1]  {\mbox{$\langle #1 \rangle$}}
\newcommand{\keyw}[1] {{\bf #1}}

\newcommand{\setproofwidth}{
\newlength{\proofwidth}
\addtolength{\proofwidth}{\textwidth}
\addtolength{\proofwidth}{-15mm}
}

\newcommand{\proofarray}[2]{
\vspace{\topsep}
\begin{minipage}[b]{\proofwidth}
\[
\begin{array}{#1}
#2
\end{array}
\]
\end{minipage}
\QED
}

\newcommand{\sidearray}[2]{
\begin{minipage}{2.8in}
\vspace{1mm}
\vspace{-\abovedisplayskip}
\begin{eqnarray*}
#1
\end{eqnarray*}
\vspace{-\abovedisplayskip}
\end{minipage}
\begin{minipage}{2.8in}
\vspace{1mm}
\vspace{-\abovedisplayskip}
\begin{eqnarray*}
#2
\end{eqnarray*}
\vspace{-\abovedisplayskip}
\end{minipage}
}

\newenvironment{ruleset}{$ \begin{array}{ll}}{\end{array} \vspace{3mm}$}

\newcommand{\rulenoprem}[1]{
#1}

\newcommand{\ruleoneprem}[2]{
\begin{array}{c}
#1 \\ \hline
#2
\end{array}}

\newcommand{\ruletwoprem}[3]{
\begin{array}{c}
#1 \hspace{4mm} #2\\ \hline
#3
\end{array}}

\newcommand{\rulethreeprem}[4]{
\begin{array}{c}
#1 \hspace{4mm} #2 \hspace{4mm} #3\\ \hline
#4
\end{array}}

\newcommand{\rulefourprem}[5]{
\begin{array}{c}
#1 \hspace{4mm} #2 \hspace{4mm} #3 \hspace{3mm} #4\\ \hline
#5
\end{array}}

% Process Relations
\newcommand{\WSimul}{\leq}
\newcommand{\CSimul}[1]{<_{{\bf C}_{#1}}}
\newcommand{\Ready}{\preceq}

% Miscellany
\newcommand{\Preserves}[2]{\mbox{$#1$ preserves $#2$}}
\newcommand{\PreSpace}{\vspace{-4mm}}
\newcommand{\PostSpace}{\vspace{-6mm}}
\newenvironment{proof}{\Proof}{\QED}
\newcommand{\SE} {{\cal E}}
\newcommand{\SF} {{\cal F}}
\newcommand{\CycProp}[1]{{\mit Cycle}(#1)}

\begin{abstract}
Model checking has found a role in the engineering of
reactive systems.  However, model checkers are still
strongly limited by the size of the system description they
can check.  Here we present a technique in which a system is
simplified prior to model checking by the application of
abstraction rules.  The rules can greatly reduce the state
space of a system description and help in understanding why
a system satisfies a property.  We illustrate the use of the
technique on examples, including Dekker's mutual exclusion
algorithm.
\end{abstract}

\section{Motivation}

Model checkers have become an important tool by which
engineers can check properties of their systems.  Model
checkers have become successful partly because they require
no user interaction.  One must only describe the system of
interest and the property to be checked.  On the other hand,
they provide little insight.  With a model checker one can
learn that a system satisfies a property, but not why it
satisfies a property.  Furthermore, the application of model
checkers is limited strongly by the number of states in the
system and in the complexity of the property to be checked.

Here we explore a technique by which an engineer can
simplify a system description by removing features of the
description believed to be irrelevant to the property to be
checked.  Simplification is performed by applying {\em
abstraction rules}.  For example, one rule allows any shared
variable to be removed.  These rules are sound in the sense
that if the simplified description satisfies a property,
then the original one does too.  However, the simplified
description may have a much smaller state space and
therefore be able to be model checked, even if the original
description could not.

This approach is also helpful in confirming one's
understanding of {\em how} a system works.  If the
simplified description satisfies a property, then we know
not only that the complete one does too, but also that our
intuition is correct.  Put another way, with our abstraction
rules one can obtain a simplified description that does not
satisfy a property even though the property {\em is}
satisfied by the original description.

Our technique is sketched in the following section of the
paper.  Then we present abstraction rules and show how they
can be applied to Dekker's mutual exclusion algorithm.  We
also discuss the use of our technique to other examples.  We
conclude with discussion and related work.

\section{Abstraction with Preorders}

Here we sketch our approach to simplifying the description
of a system relative to a property to be checked.  First of
all, we describe systems as expressions in a process
algebra.  Such {\em processes} can model both normal
terminating programs and reactive systems, which engage in
ongoing interaction with their environment.  We let $E$,
$E'$, $F$, $\ldots$ stand for processes.

We describe properties of systems as temporal logic
formulas.  In temporal logic we can express generic system
properties like deadlock freedom as well as
application-specific properties.  We let $\phi$ and $\psi$
range over temporal logic formulas and write $E \Sat \phi$
to mean that process $E$ satisfies formula $\phi$, which
captures the idea that the system modelled by $E$ has the
property expressed by $\phi$.

We want to simplify processes with respect to temporal logic
formulas.  We take a simplified process as one that stands
in a certain relation to the original process.  The kind of
process relations we are interested in are {\em preorders},
which are reflexive and transitive, but not necessarily
antisymmetric.  Suppose $\WSimul$ is a process preorder.
Then we say process $E'$ is an {\em abstract} version of
process $E$ if
\[
   E \WSimul E'.
\]
In this paper we explore preorders in which the abstract
process $E'$ can do whatever $E$ can do and possibly more.
So abstracting a system can be understood as a kind of
``loosening'' of its behavior.

We want to show that if an abstract process has a property,
then so does the original one.  To do so we find a class of
formulas such that every formula $\phi$ in the class
satisfies this logical condition:
\[
   (E \WSimul E' \mbox{ and } E' \Sat \phi) \Implies E \Sat \phi.
\]
The condition reads ``if $E'$ is an abstract version of $E$, and
$E'$ satisfies property $\phi$, then so does $E$.''

To make abstraction convenient, we do not want to have to
prove that $E \WSimul E'$ whenever a simplification to $E$
is made.  If $E$ describes a complex system this may be hard
to do.  Instead, we use a set of predefined {\em rules} for
abstraction, each of which has the form $F \WSimul F'$.  To
keep the set of rules small, each rule must be able to be
applied anywhere within a process.  For example, suppose $E$
has the form $E_1 \mid E_2$, where $E_1$ and $E_2$ are
parallel components.  Our rules should satisfy this
algebraic condition:
\[
   E_1 \WSimul E'_1 \Implies (E_1 \mid E_2 \WSimul E'_1 \mid E_2)
\]
This condition reads ``if $E_1'$ is an abstract version of $E_1$,
then $E_1' \mid E_2$ is an abstract version of $E_1 \mid E_2$.''
The preorders we use for abstraction satisfy this conditions not
just for the parallel composition operator, but for all operators
of the process notation.

\section{Processes and Properties}

We use CCS \cite{ccs} processes to describe systems.  Processes
perform actions, which are either {\em names} ($a,b,\ldots$), {\em
co-names} ($\overline{a}, \overline{b},\ldots$), or the action $\tau$.
Names and co-names satisfy $\overline{\overline{\Na{a}}} = \Na{a}$.
The set of all actions is denoted \Actions.  Process expressions have
the following syntax, where $\alpha$ ranges over actions, $L$ ranges
over sets of non-$\tau$ actions, $A$ ranges over process constants,
and $f$ ranges over relabelling functions (functions from
\Actions\ to  \Actions\ satisfying $f(\tau) = \tau$ and $f(\Co{a}) =
\overline{f(a)}$):
\[
   E ::= A              \BnfBar
         0              \BnfBar 
         \alpha.E       \BnfBar 
         E_1 + E_2      \BnfBar 
         E_1 \mid E_2   \BnfBar
         E \rest{L}     \BnfBar 
         E[f]
\]

Additionally, families of process definitions are allowed,
having the form $\{A_i \Def E_i \mid i \in I\}$.  The
meaning of CCS processes is given as a labelled transition
system in which processes are states and transitions are
labelled with actions.  If there is a transition from
process $E$ to process $F$ labelled with action $\alpha$, we
write $\trans{E}{a}{F}$ and say ``$E$ performs $\alpha$ and
becomes $F$''.  We will only briefly recall the meaning of
the CCS operators here.  Process $0$ is the deadlocked
process.  Operator $.$ is action prefixing.  Operator $+$ is
choice.  Operator $\mid$ is concurrent composition, with
synchronization between complementary actions resulting in a
$\tau$ action.  Operator $\rest{L}$ is restriction to labels
in $L$ and their complements, written $\overline{L}$.
Operator $[f]$ is relabelling by $f$.

We will additionally use a {\em hiding} operator $\hide{L}$.
We define it directly for simplicity, but it could be
defined as a derived CCS operator.
\begin{center}
    $\ruleoneprem{\trans{E}{\alpha}{E'}}
                 {\trans{E\hide{L}}{\alpha}{E'\hide{L}}}
    \hspace{2mm}
    \alpha \not \in L \cup \overline{L}$
    \hspace{5mm}
    $\ruleoneprem{\trans{E}{\alpha}{E'}}
                 {\trans{E\hide{L}}{\tau}{E'\hide{L}}}
    \hspace{2mm}
    \alpha \in L \cup \overline{L}$
\end{center}

We use a slightly extended modal mu-calculus \cite{mu-calc,mu-intro}
to express properties of processes.  We present the logic in its
positive normal form.  Formulas have the following syntax, where $L$
ranges over sets of actions and $Z$ ranges over variables:
\[
    \phi ::= Z                      \BnfBar 
             \phi _1 \MyAnd \phi _2   \BnfBar 
             \phi _1 \Or \phi _2    \BnfBar 
             \must{L}\phi           \BnfBar
             \may{L}\phi            \BnfBar
             \nu Z.\phi             \BnfBar
             \mu Z.\phi
\]

The interesting operators are the modal and fixed-point operators.
Informally, $E$ satisfies $\must{L}\phi$ (resp. $\may{L}\phi$) if all
(resp. some) $E'$ that can be reached from $E$ through an
$a$-transition ($a \in L$) satisfy $\phi$.  The fixed-point operators
bind free occurrences of $Z$ in $\phi$.  Informally, $E$ satisfies
$\nu Z.\phi$ (resp. $\mu Z.\phi$) if $E$ belongs to the greatest
(resp. least) solution of the recursive modal equation $Z = \phi$.  We
write $E \Sat \phi$ if $E$ satisfies the closed formula $\phi$.  We
use $\Tr$ as an abbreviation for the (true) formula $\nu Z.Z$, and
$\must{-L}\phi$ as an abbreviation for $\must{\Actions - L}\phi$.

We write the set of all closed modal mu-calculus formulas as $\Mu$.

\section{Weak Simulation}
\label{weak-sim}

Intuitively, one process simulates another if it can match
any action the other can do, and can continue to match in
this way indefinitely.  Here we define a particular kind of
simulation relation on processes called the {\em weak
simulation} relation and show that it has the logical and
algebraic properties we need.

The term {\em weak} indicates that the relation is based on
what we can observe of a process.  Recall that in CCS the
$\tau$ action represents internal activity of a system that
cannot be observed.  So to formalize what it means to be
observable we define a transition relation on processes
in which $\tau$ transitions do not occur.  However, 
transitions can be labelled by the new symbol $\eps$, which
represents the occurrence of zero or more $\tau$ actions.
\begin{eqnarray*}
\wtrans{E}{\eps}{F} & \Def & E (\arrow{\tau})^\ast F \\
\wtrans{E}{a}{F}	& \Def & E \warrow{\eps} \circ \arrow{a} \circ \warrow{\eps} F
                               \hspace{7mm}\mbox{($a \not = \tau$)}
\end{eqnarray*}
We write $\Actions_{obs}$ for the set $\Actions \setminus
\{\tau\} \cup \{\eps\}$ of observable actions.

\begin{definition}
A binary relation ${\cal R}$ on processes is a {\em weak simulation} if
$(E,F)$ in ${\cal R}$ implies, for all $\alpha$ in $\Actions_{obs}$:
\begin{itemize}
\item[] Whenever $\wtrans{E}{\alpha}{E'}$, then $\wtrans{F}{\alpha}{F'}$ 
        for some $F'$ such that $(E',F') \in {\cal R}$.
\end{itemize}

\noindent
$E$ {\em is weakly simulated by} $F$, written $E \WSimul F$, if
$(E,F)$ belongs to some weak simulation ${\cal R}$.
\end{definition}

If a process $E$ is weakly simulated by another process $F$,
then any sequence of actions that $E$ can perform can be
matched by $F$.  However, the converse does not hold.  For
example, $B \Def a.b.0 + a.c.0$ can match every sequence of
actions that $A \Def a.(b.0 + c.0)$ can perform, but $B$
does not weakly simulate $A$.

The weakly simulates relation is preserved by all CCS
operators, including recursive definition.  
\begin{theorem}
\label{ccs-ops-preserve-simulation}
$\WSimul$ is preserved by all CCS operators.
\end{theorem}
(Proofs for the theorems and abstraction rules in the paper
can be found in \cite{bruns:thesis}.)

To describe the properties that are preserved by the weak
simulation relation, we define a modal operator $\wmust{\,}$
that is based on the weak transition relation.  A process
$E$ satisfies formula $\wmust{a}\phi$ if all $E'$ that
can be reached from $E$ through a weak $a$-transition
satisfy $\phi$.  

We write $\Mubw$ for the set of closed modal mu-calculus
formulas containing no modal operators except $\wmust{\,}$.
An example safety property that can be expressed in $\Mubw$
is that no $a$ action can ever occur.
An example liveness property that can be expressed in
$\Mubw$ is that $a$ must eventually occur if system operation
never terminates.

If a process $F$ satisfies a formula of $\Mubw$, and $F$ weakly
simulates $E$, then $E$ also satisfies the formula.
\begin{theorem}
  Let $\phi$ be a formula of $\Mubw$.  Then
\[
  (E \WSimul E' \mbox{ and } E' \Sat \phi) \Implies E \Sat \phi.
\]
\end{theorem}

\section{Abstraction Rules}
\label{abst-ops}

We now present abstraction rules for $\WSimul$.  Each rule
has the form $E \WSimul E'$, where $E'$ is understood as the
abstract version of $E$.  

\begin{proposition}[Restriction rules]
\begin{eqnarray}
    E \rest{L}  & \WSimul & E \hide{K}\rest{L} 
                  \hspace{5mm} \mbox{if } K \subseteq L \cup \overline{L} \label{rest1} \\
    E \rest{L}  & \WSimul & E [f]\rest{L}      
                  \hspace{5mm} \mbox{if } f(\alpha) = \alpha 
                  \mbox{ for }\alpha \not \in L \cup \overline{L} \label{rest2}
\end{eqnarray}
\end{proposition}

These rules are especially important because they allow
actions that are used for process synchronization to be
hidden or renamed.  Doing so loosens synchronization between
components, which intuitively increases the number of
possible system states.  However, the rule can allow the
structure of components to be made more regular, which in
terms reduces the system state space.

\begin{proposition}[Hiding rules]
\begin{eqnarray}
(\alpha.E)\hide{L} & \WSimul & \tau.(E\hide{L}) 
                     \hspace{7mm} \mbox{ if } \alpha \in L \label{hide1} \\
(E + F)\hide{L}    & \WSimul & E\hide{L} + F\hide{L}         \label{hide2} \\
(E \mid F)\hide{L} & \WSimul & E \hide{L} \mid F \hide{L}
                     \hspace{7mm} \mbox{if } L = \overline{L} \label{hide3} 
\end{eqnarray}
\end{proposition}

\begin{proposition}[Relabelling rules]
\begin{eqnarray}
(\alpha.E)[f] & \WSimul & f(\alpha).(E[f]) \label{relab1} \\
(E + F)[f]    & \WSimul & E[f] + F[f] \label{relab2} \\
(E \mid F)[f] & \WSimul & E [f] \mid F [f] \label{relab3}
\end{eqnarray}
\end{proposition}

\begin{proposition}[Family rules]
\begin{eqnarray}
\Fam{F}                 & \WSimul & \Merge{A_j}{A_k}{\Fam{F}} \label{fam1} \\
\Fam{F}\{A \Def A + E\} & \WSimul & \Fam{F}\{A \Def E\} \label{fam2}
\end{eqnarray}
\end{proposition}

Rule \ref{fam1} allows two constants in a family of process
definitions to be ``merged''.  To merge two constants one
redefines the first of the constants by adding the
definition of the second as a choice, and then renames the
second constant to the first.  For example,
$\Merge{A}{B}{\{A \Def a.B,B \Def b.A\}}$ yields $\{A \Def
a.A + b.A\}$.  In terms of process behavior, this operation
can be understood as the combining to two process states.
The rule is used when the difference between two process
states is not important relative to the property to be
proved.

\begin{proposition}[Basic rules]
\begin{eqnarray}
E \mid 0 & \WSimul & E \label{basic1} \\
\tau.E   & \WSimul & E \label{basic4} \\
A        & \WSimul & E \hspace{7mm} \mbox{if } A \Def E \label{basic2} \\
E        & \WSimul & A \hspace{7mm} \mbox{if } A \Def E \label{basic3}
\end{eqnarray}
\end{proposition}

The abstraction rules above are typically applied in certain
combinations.  The following rules, which can be derived
from the preceding ones, allow abstraction to proceed in
coarser steps.

\begin{proposition}[Derived rules]
Let $K$ and $L$ be sets of actions such that $K = \overline{K}
\subseteq L \cup \overline{L}$, and let $f$ be a relabelling
function satisfying $f(\alpha) = \alpha$ ($\alpha \not \in L \cup
\overline{L}$).  Then
\begin{eqnarray}
  (E_1 \mid \cdots \mid E_n)\rest{L} & \WSimul & 
    (E_1 \hide{K} \mid \cdots \mid E_n \hide{K})\rest{L} \label{derived1}\\
  (E_1 \mid \cdots \mid E_n)\rest{L} & \WSimul & 
    (E_1 [f] \mid \cdots \mid E_n [f])\rest{L} \label{derived2}
\end{eqnarray}
\end{proposition}

\section{Example}

We illustrate our technique by abstracting Dekker's mutual exclusion
algorithm \cite{peterson:os} relative to a safety property and a
liveness property.  In each case, we show that if the abstract
algorithm satisfies the property, then so does the concrete algorithm.

The purpose of a mutual exclusion algorithm is to control access by a
set of processes to a shared resource so that at most one process has
access at any time.  A process is said to be in its {\em critical
section} when it has access to the resource.  The key safety property
of a mutual exclusion algorithm is that at most one process is in its
critical section at any time.  The key liveness property is that a
process wishing to enter its critical section will eventually be able
to do so.

We now present a two-process version of Dekker's mutual exclusion
algorithm.  The top level of the algorithm, written in a concurrent
while language, is as follows.

{\small
\begin{tabbing}
\keyw{begin} \= \\
\> \keyw{var} $b_1$,$b_2$,$k$; \\
\> $b_1$ := \keyw{false}; $b_2$ := \keyw{false}; \\
\> $k$ := 1; \\
\> $P_1$ \keyw{par} $P_2$ \\
\keyw{end};
\end{tabbing}
}

\noindent
Processes $P_1$ and $P_2$ coordinate via shared variables $b_1$,
$b_2$, and $k$.  Process $P_1$ is defined as follows; $P_2$ is
obtained from $P_1$ by interchanging 1 with 2 everywhere.  The
process does its useful work in the non-critical section.

{\small
\begin{tabbing}
\keyw{while} \keyw{true} \keyw{do} \\
\keyw{begin} \= \\
\> $\langle$ non-critical section $\rangle$; \\
\> $b_1$ := \keyw{true} \\
\> \keyw{while} \= $b_2$ \keyw{do} \\
\> \> \keyw{if} $k = 2$ \keyw{then} \= \keyw{begin} \= \\
\> \> \> \> $b_1$ := \keyw{false}; \\
\> \> \> \> \keyw{while} $k = 2$ \keyw{do} \keyw{skip}; \\
\> \> \> \> $b_1$ := \keyw{true} \\
\> \> \> \keyw{end}; \\
\> $\langle$ critical section $\rangle$; \\
\> $k$ := $2$; \\
\> $b_1$ := \keyw{false}\\
\keyw{end};
\end{tabbing}
}

Informally, the $b$ variables are ``request'' variables and the $k$
variable is a ``turn'' variable.  Variable $b_i$ is {\it true} if $P_i$ is
requesting entry to its critical section; variable $k$ is $i$ if it is
$P_i$'s turn to enter its critical section.  Only $P_i$ writes on
variable $b_i$, but both processes read $b_i$.

Translating the algorithm to CCS (from \cite{walker:mutex}) gives the
following family of definitions.  Actions $\Na{req_i}$,
$\Na{enter_i}$, and $\Na{exit_i}$ have been added to indicate requests
to enter, entrance to, and exit from the critical section by process
$i$.  A shared variable name (e.g., $b_1$) in a restriction operator
stands for the read and write actions of the variable
(e.g.,$\{\Na{b_1rt},\Co{b_1wt},\Na{b_1rf},\Co{b_1wf}\}$).  The
definitions of $\Ct{B2f}$ and $\Ct{P2}$ are omitted, since they are
symmetrical to $\Ct{B1f}$ and $\Ct{P1}$.

\setlength{\jot}{1mm}

\PreSpace
{\small
\begin{eqnarray*}
\Ct{K1} & \Def & \Co{kr1}.\Ct{K1} + \Na{kw1}.\Ct{K1} + \Na{kw2}.\Ct{K2} \\
\Ct{K2} & \Def & \Co{kr2}.\Ct{K2} + \Na{kw1}.\Ct{K1} + \Na{kw2}.\Ct{K2} \\
 & & \\
\Ct{B1f} & \Def & \Co{b_1rf}.\Ct{B1f} + \Na{b_1wf}.\Ct{B1f} + \Na{b_1wt}.\Ct{B1t} \\
\Ct{B1t} & \Def & \Co{b_1rt}.\Ct{B1t} + \Na{b_1wf}.\Ct{B1f} + \Na{b_1wt}.\Ct{B1t} \\
 & & \\
\Ct{P1}  & \Def & \Co{b_1wt}.\Na{req_1}.\Ct{P11} \\
\Ct{P11} & \Def & \Na{b_2rf}.\Ct{P13} + \Na{b_2rt}.(\Na{kr1}.\Ct{P11} + \Na{kr2}.\Co{b_1wf}.\Ct{P12}) \\
\Ct{P12} & \Def & \Na{kr1}.\Co{b_1wt}.\Ct{P11} + \Na{kr2}.\tau.\Ct{P12} \\
\Ct{P13} & \Def & \Na{enter_1}.\Na{exit_1}.\Co{kw2}.\Co{b_1wf}.\Ct{P1} \\
& & \\
\Ct{Dekker} & \Def & (\Ct{P1} \mid \Ct{P2} \mid \Ct{B1f} \mid \Ct{B2f} \mid K1)
                     \rest{\{\Na{b_1},\Na{b_2},\Na{k}\}}
\end{eqnarray*}}
\PostSpace
\vspace{-2mm}

\subsection{Safety}

The property of mutual exclusion is that at most one process can be in
its critical section at any time.  The property can be expressed as
formula $\Ct{ME}$ of $\Mubw$ requiring that $\Na{enter}$ and $\Na{exit}$
actions alternate:
\begin{eqnarray*}
{\mit Cycle}({L_1,L_2}) & \Def & 
    \nu X_1.\wmust{L_2}\Fa \MyAnd \wmust{- L_1,L_2}X_1 \MyAnd \wmust{L_1} \\
& & \nu X_2.\wmust{L_1}\Fa \MyAnd \wmust{- L_1,L_2}X_2 \MyAnd \wmust{L_2}X_1
\\
\Ct{ME} & \Def & {\mit Cycle}(\{\Na{enter}_1,\Na{enter}_2\},\{\Na{exit}_1,\Na{exit}_2\})
\end{eqnarray*}

Dekker's algorithm satisfies the property of mutual
exclusion because a process sets its request variable to
{\em true} before attempting to enter its critical section,
and waits for the request variable of the other process to
be {\it false} before actually entering it.  Dekker's
algorithm can be abstracted greatly with respect to mutual
exclusion because much of the algorithm's design deals with
the problem of ensuring liveness.

We now apply abstraction rules to remove other details from
the algorithm.  To make the presentation concise, we will
redefine the constants of the algorithm at each step, rather
than use new ones, and will present only the constant
definitions that were affected by the abstraction step.  As
a preliminary step we remove the indices of the
$\Na{enter}_i$ and $\Na{exit}_i$ actions by relabelling, and
hide the $\Na{req}_i$ actions, yielding process $Dekker_1$
below.

\PreSpace
{\small
\begin{eqnarray*}
\Ct{P1}  & \Def & \Co{b_1wt}.\Ct{P11} \\
\Ct{P11} & \Def & \Na{b_2rf}.\Ct{P13} + \Na{b_2rt}.(\Na{kr1}.\Ct{P11} + \Na{kr2}.\Co{b_1wf}.\Ct{P12}) \\
\Ct{P12} & \Def & \Na{kr1}.\Co{b_1wt}.\Ct{P11} + \Na{kr2}.\tau.\Ct{P12} \\
\Ct{P13} & \Def & \Na{enter}.\Na{exit}.\Co{kw2}.\Co{b_1wf}.\Ct{P1}
\end{eqnarray*}}
\PostSpace

\noindent
By the correspondence rule of \cite{action-abst} we have
\[
\Ct{Dekker} \Sat \Ct{ME} \IffArrow \Ct{Dekker}_1 \Sat {\mit Cycle}(\{\Na{enter}\},\{\Na{exit}\}).
\]
Formula ${\mit Cycle}(\{\Na{enter}\},\{\Na{exit}\})$ is also
a formula of $\Mubw$.  In all further abstraction steps we
use only the abstraction rules of Section \ref{abst-ops},
which all preserve the property ${\mit
Cycle}(\{\Na{enter}\},\{\Na{exit}\})$.

Our sketch of why Dekker's algorithm satisfies mutual
exclusion mentions only the request variables, not the turn variable
$k$.  We therefore hide $k$-related actions
$\{\Na{kr1},\Na{kw1},\Na{kr2},\Na{kw2}\}$ and their complements using
derived rule \ref{derived1}, and move the hiding inward using the
hiding rules.  We refer to $\Ct{Dekker}$ of the resulting family as
$\Ct{Dekker}_2$.

\PreSpace
{\small
\begin{eqnarray*}
\Ct{K1} & \Def & \tau.\Ct{K1} + \tau.\Ct{K1} + \tau.\Ct{K2} \\
\Ct{K2} & \Def & \tau.\Ct{K2} + \tau.\Ct{K1} + \tau.\Ct{K2} \\
& & \\
\Ct{P1} & \Def & \Co{b_1wt}.\Ct{P11} \\
\Ct{P11} & \Def & \Na{b_2rf}.\Ct{P13} + \Na{b_2rt}.(\tau.\Ct{P11} + \tau.\Co{b_1wf}.\Ct{P12}) \\
\Ct{P12} & \Def & \tau.\Co{b_1wt}.\Ct{P11} + \tau.\tau.\Ct{P12} \\
\Ct{P13} & \Def & \Na{enter}.\Na{exit}.\tau.\Co{b_1wf}.\Ct{P1}
\end{eqnarray*}}
\PostSpace

The actions related to variables $b_1$ and $b_2$ cannot all
be hidden, as variables $b_1$ and $b_2$ play a part in
ensuring mutual exclusion.  However, only some of the
$b$-related actions are involved.  We can hide the actions
that represent when a $b$ variable is read with value {\em
true}.  The effect is to allow $P_1$ to proceed as if $b_2$
is true whether $b_2$ is true or not.  Thus $P_1$ can elect
not to enter its critical section even if $b_2$ is {\it
false}.  Similarly $P_2$ can wait instead of entering its
critical section.

Applying derived rule \ref{derived1} with actions
$\{\Na{b_1rt},\Na{b_2rt}\}$ and their complements, and moving hiding
inward with the hiding rules gives the following.

\PreSpace
{\small
\begin{eqnarray*}
\Ct{B1f} & \Def & \Co{b_1rf}.\Ct{B1f} + \Na{b_1wf}.\Ct{B1f} + \Na{b_1wt}.\Ct{B1t} \\
\Ct{B1t} & \Def & \tau.\Ct{B1t} + \Na{b_1wf}.\Ct{B1f} + \Na{b_1wt}.\Ct{B1t} \\
& & \\
\Ct{P1}  & \Def & \Co{b_1wt}.\Ct{P11} \\
\Ct{P11} & \Def & \Na{b_2rf}.\Ct{P13} + \tau.(\tau.\Ct{P11} + \tau.\Co{b_1wf}.\Ct{P12}) \\
\Ct{P12} & \Def & \tau.\Co{b_1wt}.\Ct{P11} + \tau.\tau.\Ct{P12} \\
\Ct{P13} & \Def & \Na{enter}.\Na{exit}.\tau.\Co{b_1wf}.\Ct{P1}
\end{eqnarray*}}
\PostSpace

We now remove all $\tau$ actions by repeatedly applying
basic rule \ref{basic4}, and then apply family rule
\ref{fam2} to remove all unguarded occurrences of constants.

\PreSpace
{\small
\begin{eqnarray*}
\Ct{B1f} & \Def & \Co{b_1rf}.\Ct{B1f} + \Na{b_1wf}.\Ct{B1f} + \Na{b_1wt}.\Ct{B1t} \\
\Ct{B1t} & \Def & \Na{b_1wf}.\Ct{B1f} + \Na{b_1wt}.\Ct{B1t} \\
& & \\
\Ct{P1}  & \Def & \Co{b_1wt}.\Ct{P11} \\
\Ct{P11} & \Def & \Na{b_2rf}.\Ct{P13} + \Co{b_1wf}.\Ct{P12} \\
\Ct{P12} & \Def & \Co{b_1wt}.\Ct{P11} \\
\Ct{P13} & \Def & \Na{enter}.\Na{exit}.\Co{b_1wf}.\Ct{P1}
\end{eqnarray*}}
\PostSpace

Next we want to merge the states that are are outside the
critical section but have not yet requested entry to the
critical section.  To prepare for this we apply basic rule
\ref{basic3} to introduce constants $\Ct{P14}$ and
$\Ct{P24}$.

\PreSpace
{\small
\begin{eqnarray*}
\Ct{P1}  & \Def & \Co{b_1wt}.\Ct{P11} \\
\Ct{P11} & \Def & \Na{b_2rf}.\Ct{P13} + \Co{b_1wf}.\Ct{P12} \\
\Ct{P12} & \Def & \Co{b_1wt}.\Ct{P11} \\
\Ct{P13} & \Def & \Na{enter}.\Na{exit}.\Ct{P14} \\
\Ct{P14} & \Def & \Co{b_1wf}.\Ct{P1}
\end{eqnarray*}}
\PostSpace

Family rule \ref{fam1} is now applied to merge constants $\Ct{P1}$,
$\Ct{P12}$, and $\Ct{P14}$.  Similarly, constants $\Ct{P2}$,
$\Ct{P22}$, and $\Ct{P24}$ of $\Ct{P2}$ are merged.  Constants
$\Ct{K1}$ and $\Ct{K2}$ are then removed using basic rule
\ref{basic1} to obtain the following family.  We refer to
$\Ct{Dekker}$ of this family as $\Ct{Dekker}_3$.

\PreSpace
{\small
\begin{eqnarray*}
\Ct{B1f} & \Def & \Co{b_1rf}.\Ct{B1f} + \Na{b_1wf}.\Ct{B1f} + \Na{b_1wt}.\Ct{B1t} \\
\Ct{B1t} & \Def & \Na{b_1wf}.\Ct{B1f} + \Na{b_1wt}.\Ct{B1t} \\
& & \\
\Ct{P1}  & \Def & \Co{b_1wt}.\Ct{P11} + \Co{b_1wf}.\Ct{P1}\\
\Ct{P11} & \Def & \Na{b_2rf}.\Ct{P13} + \Co{b_1wf}.\Ct{P1} \\
\Ct{P13} & \Def & \Na{enter}.\Na{exit}.\Ct{P1} \\
& & \\
\Ct{Dekker} & \Def & (\Ct{P1} \mid \Ct{P2} \mid \Ct{B1f} \mid \Ct{B2f})
                       \rest{\{\Na{b_1},\Na{b_2}\}} 
\end{eqnarray*}}
\PostSpace

The abstract algorithm contains little more than the
protocol for handling the request variables to ensure mutual
exclusion.  While process \Ct{Dekker} has 153 states,
$\Ct{Dekker}_3$ has only 16 states.  With a tool like the
Concurrency Workbench \cite{cwb} it is easy to show that
$\Ct{Dekker}_3$ satisfies ${\mit
Cycle}(\{\Na{enter}\},\{\Na{exit}\})$.  Since
$\Ct{Dekker}_3$ was obtained by abstraction rules that
preserve ${\mit Cycle}(\{\Na{enter}\},\{\Na{exit}\})$ we
know $\Ct{Dekker}_1$ also satisfies
$\CycProp{\{\Na{enter}\},\{\Na{exit}\}}$.  Then, since
\[
  \Ct{Dekker}_1 \Sat \CycProp{\{\Na{enter}\},\{\Na{exit}\}}
  \Implies 
  \Ct{Dekker} \Sat \Ct{ME}  
\]
we know $\Ct{Dekker} \Sat \Ct{ME}$.

The soundness of our abstraction of Dekker's algorithm does
not depend on the assumptions we made about the shared
variables in it.  For example, our decision to hide actions
of variable $k$ was based on the assumption that $k$ plays
no part in ensuring mutual exclusion.  Regardless of the
truth of this assumption, the algorithm will satisfy the
formula expressing mutual exclusion if the abstract version
of it does.  Since mutual exclusion does hold in the
abstract algorithm, we know not only that it also holds in
the concrete algorithm, but also that our assumption about
variable $k$'s role in mutual exclusion is correct.

\subsection{Liveness}

An important liveness property for mutual exclusion
algorithms is that if one process requests entry to its
critical section, then it will not wait forever while the
other processes continue to enter their critical sections.
Here we use our abstraction rules to show that Dekker's
algorithm satisfies this property, with process 2 as
the requesting process.  

Dekker's algorithm satisfies this property only under the
fairness assumption that the requesting process continues to
execute after it requests entry to its critical section.  We
handle the fairness assumption by incorporating it into the
formula we use to express the liveness property.  However,
we must also revise process 2 by adding ``probe'' action
$\Na{p2}$, which continues to occur while process $P_2$ is
waiting to enter its critical section.

\PreSpace
{\small
\begin{eqnarray*}
\Ct{P2}  & \Def & \Co{b_2wt}.\Na{req_2}.\Ct{P21} \\
\Ct{P21} & \Def & \Na{b_1rf}.\Ct{P23} + \Na{b_1rt}.\Na{p2}.(\Na{kr2}.\Ct{P21} + \Na{kr1}.\Co{b_2wf}.\Ct{P22}) \\
\Ct{P22} & \Def & \Na{kr2}.\Co{b_2wt}.\Ct{P21} + \Na{kr1}.\Na{p2}.\tau.\Ct{P22} \\
\Ct{P23} & \Def & \Na{enter_2}.\Na{exit_2}.\Co{kw1}.\Co{b_2wf}.\Ct{P2}
\end{eqnarray*}}
\PostSpace

The following $\Mubw$ formula expresses that process 2 will
not wait forever to enter its critical section while process
1 continues to enter its critical section, provided that
process 2 continues to execute while waiting.
\begin{tabbing}
\hspace{4mm}   $\Ct{Live} \Def \nu X$.\=$\wmust{-\Na{req}_2}X \MyAnd$\\
 \>$\wmust{\Na{req}_2} \mu Y$.\=$\nu X_1.\wmust{-\Na{enter}_1,\Na{enter}_2,\Na{p2}}X_1 \MyAnd \wmust{\Na{enter}_2}X \MyAnd \wmust{\Na{enter}_1}$\\
 \> \> $\nu X_2.\wmust{-\Na{enter}_1,\Na{enter}_2,\Na{p2}
}X_2 \MyAnd \wmust{\Na{enter}_2}X \MyAnd \wmust{\Na{p2}}Y$
\end{tabbing}
Paraphrasing the formula gives: ``after $\Na{req}_2$ occurs,
no path containing infinitely many $\Na{enter}_1$ and
$\Na{p2}$ actions but no $\Na{enter}_2$ actions, can
occur''.  Informally, Dekker's algorithm satisfies this
property because Process 1 will only enter its critical
section if process 2's request variable is false, and will
set the turn variable to 2 upon leaving its critical
section.  If Process 2 has requested entry but not yet
entered its critical section, and the turn variable is set
to 2, then it will eventually set its request variable to
{\em true} and keep it {\em true} until after it exits its
critical section.

To abstract Dekker's algorithm relative to this formula, we
first hide both $\Na{exit}$ actions and action $\Na{req}_1$.
This step is justified by the correspondence rule of
\cite{action-abst}.  Then we hide all actions corresponding
to reads of $b$ variables of value {\em true}, to all reads
and writes of $b_1$ variables, and to reads of $k$ of value
2.  Finally all $\tau$ actions are removed and some
constants are merged.  The abstract form of the algorithm is
as follows.

\PreSpace
{\small
\begin{eqnarray*}
\Ct{B2f} & \Def & \Co{b_2rf}.\Ct{B2f} + \Na{b_2wf}.\Ct{B2f} + \Na{b_2wt}.\Ct{B2t}\\
\Ct{B2t} & \Def & \Na{b_2wf}.\Ct{B2f} + \Na{b_2wt}.\Ct{B2t}\\
& & \\
\Ct{K1} & \Def & \Co{kr1}.\Ct{K1} + \Na{kw1}.\Ct{K1} + \Na{kw2}.\Ct{K2}\\
\Ct{K2} & \Def & \Na{kw1}.\Ct{K1} + \Na{kw2}.\Ct{K2}
\end{eqnarray*}
\sidearray{
\Ct{P1}  & \Def & \Na{b_2rf}.\Ct{P13} + \Na{kr1}.\Ct{P1}\\
\Ct{P13} & \Def & \Na{enter}_1.\Co{kw2}.\Ct{P1}
}{
\Ct{P2}  & \Def & \Co{b_2wt}.\Na{req}_2.\Ct{P21}\\
\Ct{P21} & \Def & \Ct{P23} + \Na{p}_2.\Ct{P21} + \Na{kr1}.\Co{b_2wf}.\Ct{P22}\\
\Ct{P22} & \Def & \Co{b_2wt}.\Ct{P21} + \Na{kr1}.\Na{p}_2.\Ct{P22}\\
\Ct{P23} & \Def & \Na{enter}_2.\Co{kw1}.\Co{b_2wf}.\Ct{P2}
}
\begin{eqnarray*}
\Ct{Dekker} & \Def & (\Ct{P1} \mid \Ct{P2} \mid \Ct{B2f} \mid \Ct{K1})\rest{\{\Na{b_2},\Na{k}\}}
\end{eqnarray*}}
\PostSpace
\vspace{-3mm}

\section{Other Examples}

Dekker's algorithm was presented as our main example because
it is a non-trivial algorithm but simple enough to work
through in detail.  We have used our abstraction rules on
other examples, including other the mutual exclusion
algorithms of Dijkstra \cite{dijkstra:mutex}, Knuth
\cite{knuth:mutex}, and Peterson \cite{peterson:os}.  For
example, using our abstraction operations we reduced a three
process version of Dijkstra's algorithm from 10570 to 109
states.  In abstracting two-process versions of the
algorithms we found that all the algorithms reduce to
virtually the same 16 state algorithm.  Thus, abstraction
shows that the same idea is used to achieve safety in all
the algorithms.  However, in abstracting the processes with
respect to liveness, different algorithms are reached.

We have used our abstraction technique with preorders other
than weak simulation.  In \cite{bruns:thesis} we use the
{\em anonymous ready simulation} preorder to abstract {\em
prioritized} CCS processes \cite{CH:priorities}.  The sole
abstraction rule here is the adding of priorities to a
process.  Using this rule we have been able to check a
property of Ben-Ari's concurrent garbage collection
algorithm \cite{ben-ari:garbage-collection}, which we could
not do without abstraction.

\section{Discussion and Related Work}

We have presented a general technique for abstracting a
system description relative to a property to be checked, and
illustrated it by developing abstraction rules for CCS
processes.  We would like to develop similar abstraction
rules for notations that are more suitable for the
description of complex systems.  One possibility is {\em
value-passing CCS} \cite{ccs}, which we used in our work in abstraction
using priorities.  Other possibilities include concurrent
while languages, LOTOS \cite{lotos}, CRL \cite{mu-crl}, or
even a programming language such as C.  The main problem
faced in using a higher-level notation is to establish the
algebraic conditions we need to apply the technique.  One way
to deal with this problem is to try to define the notation
in terms of a low-level notation like CCS.  Then if the
algebraic condition we need holds of CCS, it also holds of
the higher-level notation.

The possibilities for automation with our technique are not
clear.  In the example we presented the abstraction rules
were selected and applied manually.  It would not be
difficult to provide tool support for the automatic
application of rules.  A more interesting question is
whether it would be worthwhile to try to select abstraction
rules automatically.  A problem is that there may be many
abstracted forms of an algorithm that have small state
spaces but fail to satisfy the property of interest.

There is existing work on abstraction using preorders, but
none that gives abstraction rules.  Lynch
\cite{lynch:possibility-mappings} abstracts I/O automata
using the simulation preorder, but requires that a
simulation relation be invented and checked.  The logical
effect of abstraction with simulation is not given; it is
only stated that simulation is a stronger relation than
trace inclusion.  Clarke {\it et al}
\cite{clarke:checking-abst} abstract finite-state,
procedural programs by mappings on program inputs and
outputs.  No abstraction operations on control structure are
given.  The technique is justified by showing that a kind of
homomorphic mapping on transition systems preserves
CTL$^\ast$ formulas with only the universal path quantifier.
This mapping is stronger than the simulation relation, and
permits only operations that immediately reduce the state
space.  Cleaveland and Riely \cite{cleaveland:abst} abstract
value-passing CCS processes by mappings on data domains.
They show that an abstract value-passing process is greater
in the specification preorder than the original process.
The abstraction operations do not operate on process
structure, and their logical effects are not given.

\bibliography{thesis,safety}

\begin{thebibliography}{10}

\bibitem{ben-ari:garbage-collection}
Mordechai Ben-Ari.
\newblock Algorithms for on-the-fly garbage collection.
\newblock {\em ACM Transactions on Programming Languages and Systems},
  3(6):333--344, 1984.

\bibitem{lotos}
T.~Bolognesi and E.~Brinksma.
\newblock Introduction to the specification language {LOTOS}.
\newblock In van Eijk, Vissars, and Diaz, editors, {\em The Formal Description
  Technique LOTOS}. Elsevier, 1989.

\bibitem{action-abst}
Glenn Bruns.
\newblock A practical technique for process abstraction.
\newblock In {\em Proceedings of CONCUR '93, LNCS 715}, pages 37--49, 1993.

\bibitem{bruns:thesis}
Glenn Bruns.
\newblock {\em Process Abstraction in the Verification of Temporal Properties}.
\newblock PhD thesis, University of Edinburgh, 1997.
\newblock Published as report ECS-LFCS-98-380 by the Department of Computer
  Science.

\bibitem{clarke:checking-abst}
Edmund~M. Clarke, Orna Grumberg, and David~E. Long.
\newblock Model checking and abstraction.
\newblock In {\em Proceedings of the 19th Annual ACM Symposium on Principles of
  Programming Languages}, pages 343--354, 1992.

\bibitem{CH:priorities}
Rance Cleaveland and Matthew Hennessy.
\newblock Priorities in process algebra.
\newblock {\em Information and Computation}, 87(1/2), 1990.

\bibitem{cwb}
Rance Cleaveland, Joachim Parrow, and Bernhard Steffen.
\newblock The {C}oncurrency {W}orkbench: A semantics based tool for the
  verification of concurrent systems.
\newblock {\em ACM Transactions on Programming Languages and Systems},
  15(1):36--72, 1993.

\bibitem{cleaveland:abst}
Rance Cleaveland and James Riely.
\newblock Testing-based abstractions for concurrent systems.
\newblock In {\em Proceedings of CONCUR '94, LNCS 836}, pages 417--432, 1994.

\bibitem{dijkstra:mutex}
E.W. Dijkstra.
\newblock Solution of a problem in concurrent programming control.
\newblock {\em Communications of the ACM}, 8(9):569, 1965.

\bibitem{mu-crl}
J.F. Groote and A.~Ponse.
\newblock The syntax and semantics of $\mu${CRL}.
\newblock Technical Report CS-R9076, Centre for Mathematics and Computer
  Science, {CWI}, 1990.

\bibitem{knuth:mutex}
D.E. Knuth.
\newblock Additional comments on a problem in concurrent programming control.
\newblock {\em Communications of the ACM}, 9(5), 1966.

\bibitem{mu-calc}
D.~Kozen.
\newblock Results on the propositional mu-calculus.
\newblock {\em Theoretical Computer Science}, 27:333--354, 1983.

\bibitem{lynch:possibility-mappings}
Nancy~A. Lynch.
\newblock Multivalued possibilities mapping.
\newblock In J.W. de~Bakker, W.-P. de~Roever, and G.~Rozenberg, editors, {\em
  Stepwise Refinement of Distributed Systems}, pages 519--543, 1989.
\newblock LNCS 430.

\bibitem{ccs}
Robin Milner.
\newblock {\em Communication and Concurrency}.
\newblock Prentice Hall International, 1989.

\bibitem{peterson:os}
J.L. Peterson and A.~Silberschatz.
\newblock {\em Operating System Concepts}.
\newblock Addison Wesley, 1985.

\bibitem{mu-intro}
Colin Stirling.
\newblock An introduction to modal and temporal logics for {CCS}.
\newblock In A.~Yonezawa and T.~Ito, editors, {\em Concurrency: Theory,
  Language, and Architecture}, pages 2--20, 1989.
\newblock LNCS 491.

\bibitem{walker:mutex}
D.~Walker.
\newblock Automated analysis of mutual exclusion algorithms using {CCS}.
\newblock {\em Formal Aspects of Computing}, 1:273--292, 1989.

\end{thebibliography}

\end{document}